\documentstyle[12pt,aaspp4]{article}
\tighten

\begin{document}


\def\zz{\hang\noindent}

\def\kms{km s$^{-1}$}
\def\pix{pix$^{-1}$}
\def\deg{$^\circ$}

\def\mic{{$\mu$m}}

\def\h2o{H$_2$O}

\def\ak{{\it $A_K$}}

\def\teff{$T_{\rm eff}$}

\def\aple{$\mathrel{\hbox{\rlap{\hbox{\lower4pt\hbox{$\sim$}}}\hbox{$<$}}}$
}
\def\apge{$\mathrel{\hbox{\rlap{\hbox{\lower4pt\hbox{$\sim$}}}\hbox{$>$}}}$
}


\title{$H-$Band Spectroscopic Classification of OB Stars}

\author{R. D. Blum\altaffilmark{1}, T. M. Ramond, P. S. Conti}
\affil{JILA, University of Colorado\\Campus Box 440, Boulder, CO,
80309\\rblum@casa.colorado.edu\\ramond@casa.colorado.edu\\
pconti@jila.colorado.edu}

\author{D. F. Figer}
\affil{Division of Astronomy, Department of Physics \& Astronomy,
\\ University of California, Los Angeles, CA, 90095 \\
figer@astro.ucla.edu}

\author{K. Sellgren}
\affil{Department of Astronomy, The Ohio State University\\
174 W. 18th Ave., Columbus, OH, 43210\\ sellgren@payne.mps.ohio-state.edu}

\altaffiltext{1}{Hubble Fellow}

\centerline{{\it accepted for publication in the} AJ}

\begin{abstract}

We present a new spectroscopic classification for OB stars based on
$H-$band (1.5 \mic \ to 1.8 \mic) observations of a sample of stars with
optical spectral types. Our initial sample of nine stars demonstrates
that the combination of \ion{He}{1} 1.7002 \mic \
and H Brackett series absorption can be used to determine
spectral types for stars between $\sim$ O4 and B7 (to within $\sim$
$\pm$ 2 sub--types). We find that the
Brackett series exhibits luminosity effects similar to the Balmer
series for the B stars.
This classification scheme will be useful in studies of optically
obscured high mass star forming regions. In addition, we present
spectra for the OB stars near 1.1 \mic \ and 1.3 \mic \ which
may be of use in analyzing their atmospheres and winds.

\end{abstract}


\newpage
\section{INTRODUCTION}
OB stars are massive and thus short lived. Because they have short
lives, they will be confined to regions relatively close to their
birthplaces and will be found close to the Galactic plane.
The youngest massive stars will also still be in or near their dusty
star forming environment.
Due to the strong extinction from interstellar dust
at optical wavelengths, OB stars
at large distances from the sun, or in young star forming regions,
are more easily studied at near infrared (1 $-$ 5 \mic) \ or longer
wavelengths.
Note that the extinction at $V$ is about 10 times greater in $magnitudes$ than
at 2.2 \mic, and for \aple 3.5 $\lambda$ \apge 0.9 \mic, the extinction can be
represented by a power--law dependence: $A_{\lambda}$ / $A_{2.2 \mu m}$ =
$(\lambda / 2.2)^{-1.7}$ (Mathis 1990\markcite{mat90}).

The advent of sensitive infrared array detectors has made
near infrared classification schemes commonplace for a wide range of
stellar spectral types. For example, Kleinmann \& Hall
(1986)\markcite{kh86},
Greene \& Meyer (1995\markcite{gm95}), Ali et al. (1995)\markcite{ali95},
and Ram\'{\i}rez et al. (1997)\markcite{ram97}
have presented $K-$band spectra of late--type stars useful for
classification
at 2 \mic. Lan\c{c}on \& Rocca--Volmerange (1992)\markcite{lv92} have
presented $H$ and $K$ spectra of optically classified
stars ranging in spectral type from O6 to M7 (only three OB stars were
included). Eenens et al. (1991\markcite{eww91}),
Blum et al. (1995$a$\markcite{blum95}),
Tamblyn et al. (1996\markcite{t96}), Morris et al.
(1996\markcite{m96}), and Figer et al. (1996\markcite{fmm96}, 1997
\markcite{fmn97})
have presented $K-$band spectra of
known Wolf--Rayet and other emission--line stars.
A detailed classification scheme in the $K-$band for early--type
(OB) stars has been presented by
Hanson \& Conti (1994)\markcite{hc94} and Hanson et al.
(1996\markcite{hcr96}, hereafter HCR96).
All these infrared classification spectra are now routinely applied to the
study of obscured stars, typically in star forming regions and/or along
sight
lines with large interstellar obscuration by dust (e.g. Greene \& Meyer
1995\markcite{gm95}; Blum et al. 1995$a$\markcite{blum95},$b$\markcite{bsd95};
Tamblyn et al. 1996\markcite{t96}; Figer et al. 1996\markcite{fmm96};
Ali 1996\markcite{ali96b}; Blum et al. 1996\markcite{blum96}).

Classification of hot
stars at near infrared wavelengths is challenging because the
photospheric absorption features used in the classification
(primarily H, \ion{He}{1}, and \ion{He}{2}) are generally weaker
than the molecular and atomic (metal) lines used for late--type stars.
In addition, there can also be strong circumstellar emission which
dilutes the photospheric lines.
It is this latter problem that motivates the present work. Since we
expect circumstellar dust emission to dominate the $K-$band spectra
of some stars in star forming regions
(e.g., in M17, Hanson \& Conti 1995\markcite{hc95}),
a more robust
classification scheme for high mass star forming regions
will also utilize shorter wavelengths which
may be less
affected by circumstellar excess emission. Conversely, we wish to observe
at the longest possible wavelengths shortward of $K$ to overcome the
extinction of intervening dust. In this paper, we present a preliminary
classification of OB stars based on spectra between 1.5 \mic \ and 1.8
\mic \ (i.e. in the $H-$band). We also present spectra of the same
stars between 1 \mic \ and 1.3 \mic \ which may prove useful in the
analysis of massive stars atmospheres.

\section{OBSERVATIONS and DATA REDUCTION}

The OB stars were observed on the nights of 9, 10, and 13 June 1996 using
the Ohio State Infrared Imager and Spectrometer (OSIRIS) on the Perkins
1.8--m telescope of the Ohio Wesleyan and Ohio State Universities
at the Lowell Observatory. The Perkins telescope is located
on Anderson Mesa near Flagstaff, Arizona. OSIRIS is more fully
described by DePoy et al. (1993)\markcite{dep93}. The target stars
were taken from the catalog of HCR96 and are listed, along with their
optical spectral types, in Table~\ref{stars}.
In addition to the target OB stars, we observed A--type and G--type stars
for use in canceling telluric absorption features. We will refer to
these stars as ``atmospheric standards.''

\begin{table}
\dummytable\label{stars}
\end{table}

\begin{table}
\dummytable\label{eqw}
\end{table}

OSIRIS was used in cross--dispersed mode which gives
$\lambda/\Delta\lambda \approx 570$ (2 pixels) while covering the
$J$ ($\lambda _{\circ} \approx$ 1.25 \mic, $\Delta\lambda \approx$ 0.30
\mic), $H$ ($\lambda_\circ \approx$ 1.65 \mic, $\Delta\lambda \approx$
0.37 \mic),
and $K$ ($\lambda_\circ \approx$ 2.20 \mic, $\Delta\lambda \approx$ 0.50
\mic) bands simultaneously.
One star (HD183143) was also observed in long slit mode
at $J$ and $H$. Long slit mode is
similar to the cross--dispersed mode (same spectral resolution
and spatial scale) except only one wavelength band is acquired
at a time
and the long slit fills the entire spatial dimension on the array.
Similar results were obtained from both sets of spectra; however, the
long slit mode spectra are slightly higher S/N and are therefore
presented here. In addition, five of the stars observed in
cross--dispersed mode were also observed at $\lambda$ $\approx$ 1.10
\mic \ ($\Delta\lambda$ $\approx$ 0.25 \mic),
a bandpass which we call $I'$, in long slit mode.
Analysis of $H-$band OH sky lines results in linear dispersions
of 9.70, 11.60, 14.53, and 19.37
\AA \ \pix \ at $I'$, $J$, $H$, and $K$, respectively. The sky lines were
identified using the list of Oliva \& Origlia (1992)\markcite{oo92}.
$I'$, $J$, and $K$ band dispersions follow from the $H-$band dispersion
and the appropriate order number, m (m $=$ 6, 5, 4, 3 for $I'$, $J$, $H$,
$K$).
In cross--dispersed mode, OSIRIS has a $\sim$ 60$''$ $\times$ 5$''$
slit (1.5$''$ \pix). The target stars were observed at \apge 12
uniformly spaced
positions along the slit. Combining spectra over a uniform grid along
the slit greatly reduces systematic errors which may be introduced, for
example, by scattered light in the dome flats and fringing
(see Blum et al. 1995$a$\markcite{blum95} for a discussion
of these problems in OSIRIS). The seeing varied
throughout each night from 2$''$ to 4$''$. None of the nights
were photometric; we observed through thin clouds. This will not
affect our results since we are interested in the relative
intensities and line strengths, not absolute fluxes.

All basic data reduction was accomplished using
IRAF\setcounter{footnote}{1}\footnote{IRAF is distributed by the
National Optical Astronomy Observatories.}. Each image
was flat--fielded using dome flats and then sky subtracted using
another image from the grid with the star displaced by several positions
along the slit (cross--dispersed mode) or with a median combination
sky image (long slit mode). Individual spectra were extracted from each
program
star image and atmospheric standard image using IRAF ``APEXTRACT.''
Synthesized
apertures $\pm$ 3 pixels wide were used. The entire grid of 1--d spectra
for each star was then combined (after scaling). In the case of the
long slit spectra, the individual 1--d spectra were first shifted
(0 to $\sim$ 4 pixels) to account for anamorphic demagnification along the
slit spatial dimension.

The final spectra were obtained by ratioing the program stars with a
atmospheric standard star which had first been corrected for intrinsic
H absorption lines
(P$\beta$ 1.281 \mic, and Brackett series
1.51 \mic \ to 2.165 \mic).
The atmospheric standards were observed in pairs of
A--type and G--type stars. The H lines in the A stars were corrected in the
following way. The majority (seven) of our program stars were corrected
using an A0 V and a G2 V. In this case, the G2 star (BS 6847) was first
corrected for H line absorption using the NOAO solar atlas
(Livingston \& Wallace 1991\markcite{lw91}).
Next a ratio of the A star (BS 7734) to the corrected G star was made.
Since these two were
taken at nearly the same airmass, the resulting ratio contains
essentially only the H line spectrum of the A star and perhaps some
``emission'' lines due to metal lines in the G star. The H lines in this
ratio were then fit with Gaussian profiles. The resultant line fits
were used to correct the H absorption in the A star.
The two remaining stars (HD183143 and BD+24 3866) were corrected using
an A3 V (BS 7958) and a G9 III (BS 7760). We have no intrinsic spectrum
to correct BS 7760 with, so the $J$ and $K-$band lines were corrected
by eye. No correction was made for the $H-$band Brackett lines in
BS 7760.
Since BS 7760 was not corrected using a matching
intrinsic H spectrum, we are less confident in the precise H line
absorption resulting in HD183143 and BD+24 3866; however, none of
our primary results is sensitive to the accuracy of our
measurement of H absorption
in the program stars but only the appearance or absence of a
definitive Brackett series.
No telluric correction was made for the
$I'$ spectra as a cursory inspection of the
spectra shows no strong telluric lines. We did obtain A and G star
spectra for use in making a similar telluric correction as described above.
Making this correction shows no difference in the final $I'$ spectra
of the OB stars at the few percent level and serves only to introduce
more noise.

We have also included two $H-$band spectra (HD 157955 and HD 169033)
which were obtained as part of a program concentrating on $K-$band
spectra of stars in an inner Galaxy star forming region called the
Quintuplet (Figer et al.
1996\markcite{fmm96}). These spectra have similar spectral (16.75 \AA \
\pix) resolution as the OSIRIS data and spatial scale somewhat
smaller (0.7$''$ \pix); see Figer et al. (1995\markcite{fmm95}).
The telluric correction was made by ratioing the
stars to a Quintuplet star (q3 in Moneti et al. 1994\markcite{mgm94})
and a dusty Wolf-Rayet star (WR118),
neither of which have spectral lines near the \ion{He}{1} or Brackett
lines.
These stars were observed at higher airmass ($\sim$ 2)
than the other stars presented here ($\sim$ 1) and may not be corrected
as well (the higher Brackett lines appear to suffer from incomplete
correction).

\section{RESULTS and DISCUSSION}
\subsection{The $H-$band Spectra}

The final $H-$band spectra are shown in Figure~\ref{spect}.
To highlight the absorption features,
we have divided each spectrum by a low order fit to the
continuum. We do not present
our $K-$band spectra here since spectra for these stars
have been presented by HCR96 (at higher spectral resolution).
We note that our lower resolution
$K-$band spectra reproduce the basic features of the HCR96 spectra.
Line identifications used throughout this paper
were taken from Weise et al. (1966\markcite{w66},
H, \ion{He}{1}), and Garcia \& Mack (1965\markcite{gm65}, \ion{He}{2}).
All wavelengths used and/or quoted are in air.
Line equivalent widths (Table~\ref{eqw}) were measured
by fitting Gaussian profiles
using the LINER program (Pogge 1997\markcite{pog97}). Uncertainties
are the formal one sigma errors derived from the rms scatter in the
continuum corresponding to a bandpass equal to
the full width at zero intensity of the lines. The errors do not
reflect the line fit.

\subsubsection{The Brackett Series}

\begin{figure}
\plotfiddle{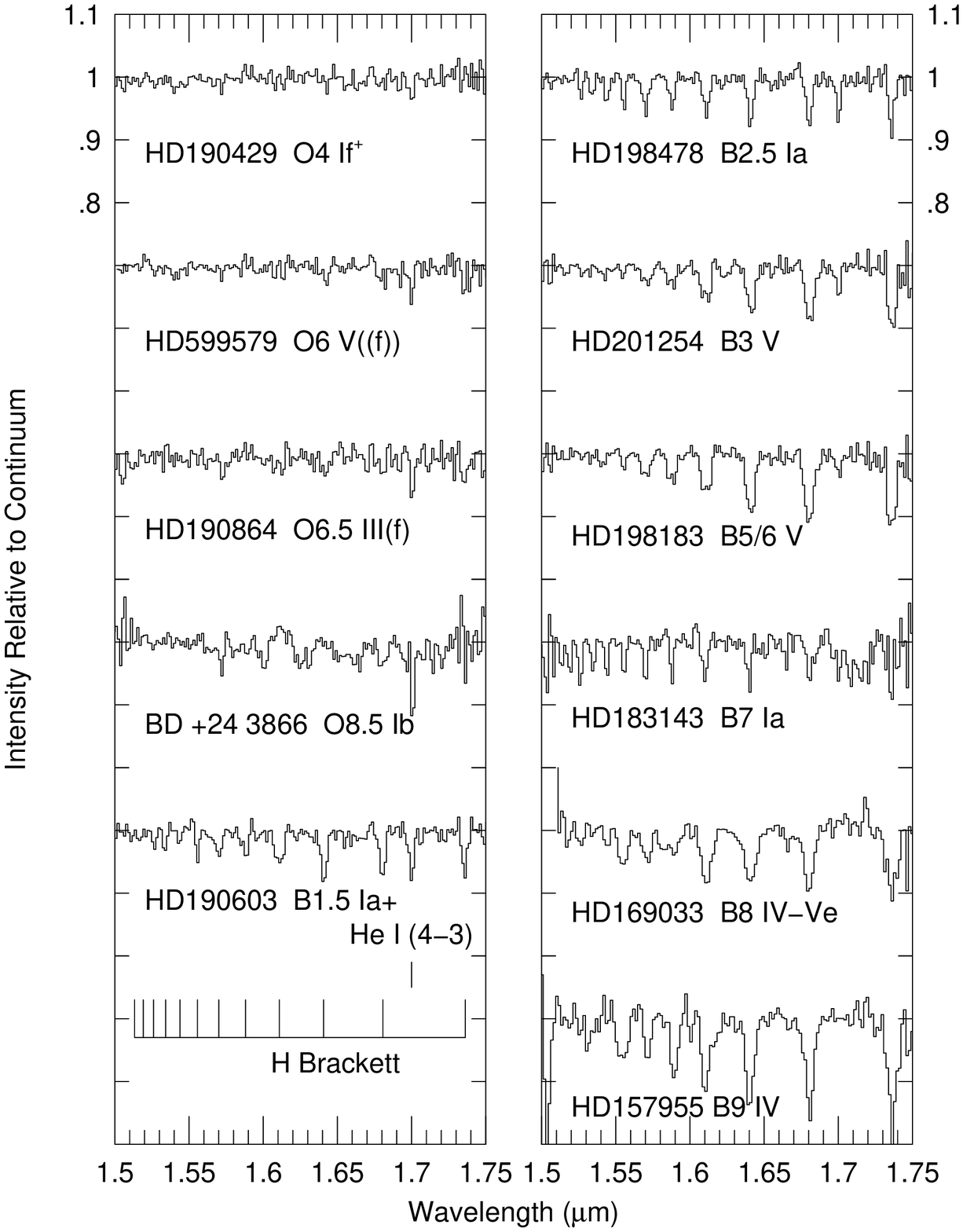}{7 in}{0}{80}{80}{-260}{-40}
\caption[]{$\lambda / \Delta\lambda \approx 570$ $H-$band
spectra of OB stars (The two latest B stars are from an independent
data set and have $\lambda / \Delta\lambda \approx$ 525; see text). The
spectra have been normalized by a low order fit to the continuum.
\ion{He}{1} 1.7002 \mic \ and H Brackett lines are the key to the present
$H-$band classification of OB stars. The first Brackett line is Br10.
See text and Figure~\ref{he1}. The intensity scale
is the same for each star.
}
\label{spect}
\end{figure}

The H lines are sensitive to the excitation in the stellar atmosphere.
For example, the strength of the
well--known Balmer lines increases with \teff \ as more atoms
are excited to the n$=$2 level. The Balmer lines
reach peak strength at $\sim$ 9000 K (Gray
1992\markcite{g92}) as Hydrogen becomes appreciably ionized at higher
temperatures.
The Brackett series excitation is similar to that
for the Balmer series (10.8 eV compared to 10.2 eV), so we expect the
lines to behave in a similar manner. This basic behavior with \teff \
(spectral type)
is seen in Figure~\ref{spect}.

Inspection of Figure~\ref{spect} suggests the Brackett lines behave
similarly to the Balmer lines in early--type stars
as first noted by Adams \& Joy (1922\markcite{aj22},
1923\markcite{aj23}): the width of the
H lines decreases with stellar luminosity
for the B stars; further, the last line visible in the series increases
with luminosity. Br19 is confidently detected in the B supergiants while
Br15 or Br16 is the last of the series for the B dwarfs.
The measured line widths (FWHM $\sim$ 60 \AA \ for Br11--Br15)
for the two B dwarf stars are
roughly two times the instrumental resolution (FWHM $=$ 29 \AA).
This can be compared to a maximum of 44 \AA \ in the B supergiants for
the same lines.
These ``luminosity'' effects
are well understood as resulting from the
linear Stark effect (Hulbert 1924\markcite{h24}; Struve 1929\markcite{s29};
Gray 1992\markcite{g92}) which is sensitive to the electron pressure.

In Figure~\ref{spect} it can be noted that the  upper Brackett lines are
missing in the earliest O-type stars, although Br$\gamma$ is present in
these same stars (HCR96). There are, as yet, no published model predictions
for the upper Brackett series lines,
but we suspect their absence in the
earliest O-type stars, while Br$\gamma$ is still present, is analogous
to the case for the Balmer lines.
LTE model predictions of Auer \&
Mihalas (1972) clearly underestimate H$\gamma$ absorption compared to
observations, while their non--LTE predictions provide a much better
fit (Conti 1973). Peterson \& Scholz (1971)\markcite{ps71} demonstrate
that the difference between the observations and LTE predictions is
much less for the higher series member H8 than for H$\gamma$.
The upper series members of the Brackett lines probably are
still close to LTE (similar to H8) and thus, quite weak due to the
advanced ionization state of hydrogen.
The lower series members,
such as Br$\gamma$, exhibit non--LTE behavior (similar to H$\gamma$)
and are thus still present.

\subsubsection{\ion{He}{1} and \ion{He}{2}}
\begin{figure}
\plotone{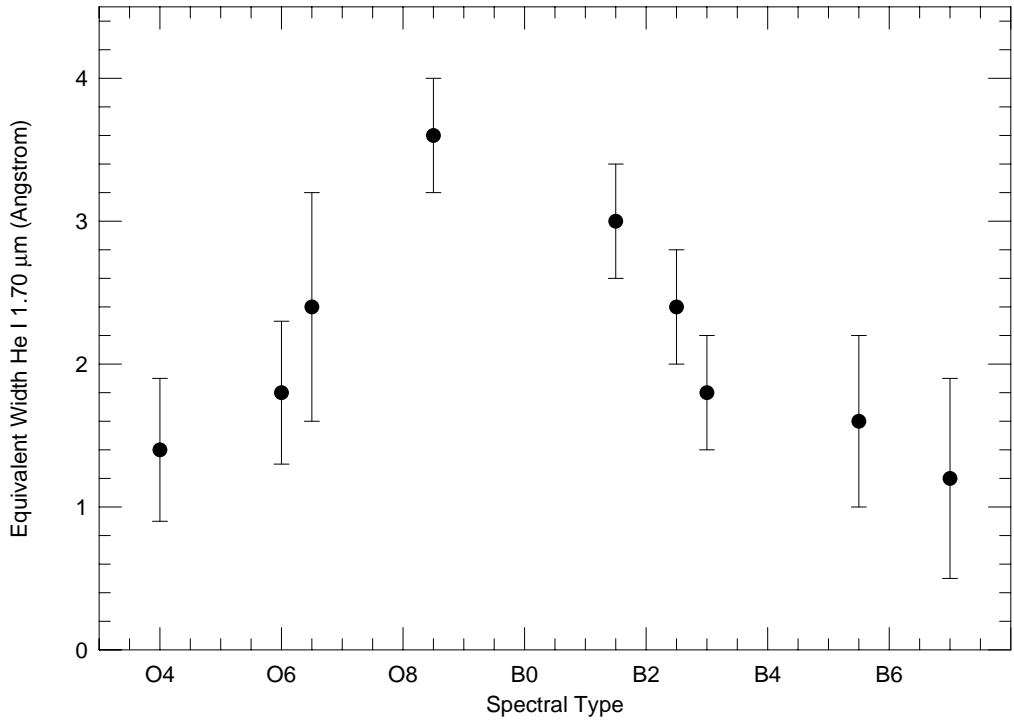}
\caption[]{\ion{He}{1} 1.7002 \mic \ equivalent width versus spectral type.
O type stars and late B type stars are distinguished
by the presence or absence of H Brackett absorption; see
Figure~\ref{spect}.
}
\label{he1}
\end{figure}

A relatively strong line of \ion{He}{1} (4--3) is visible at 1.7002 \mic.
To be useful for classification purposes, the wavelength range of
interest must contain various lines sensitive to different excitation in
the stellar atmosphere. In the present case, the Brackett series
lines (as discussed above) and \ion{He}{1} 1.7002 \mic \ line
are sensitive to lower and higher \teff, respectively, in the atmospheres
of OB stars.
In Figure~\ref{he1}, we plot the equivalent width of the \ion{He}{1}
(4--3) 1.7002 \mic \ line versus the spectral type (Table~\ref{stars}).
This line is at maximum strength in the late O/early B types.
In the earliest O stars, the Brackett series will be weak or absent
(Figure~\ref{spect}), whereas in the late B stars, the Brackett lines are
stronger. These features, therefore, provide a rough spectral class in the
$H-$band for stars in the range $\sim$ O4 to B7
(Figure~\ref{he1} suggests a spectral type could be determined
to $\sim$ $\pm$ 2 sub--types). Note that \ion{He}{1} 1.7002 \mic \ is
detected in HD183143 (B7 Ia) even though the telluric correction
appears worse for this star: compare the region just to the red wavelength side
of the 1.7 \mic \ line. The dip in the spectrum is due to a relatively poor
air--mass match between object and standard, but the He I 1.7002 \mic \ line
is visible in the uncorrected spectrum.

\begin{figure}
\plotfiddle{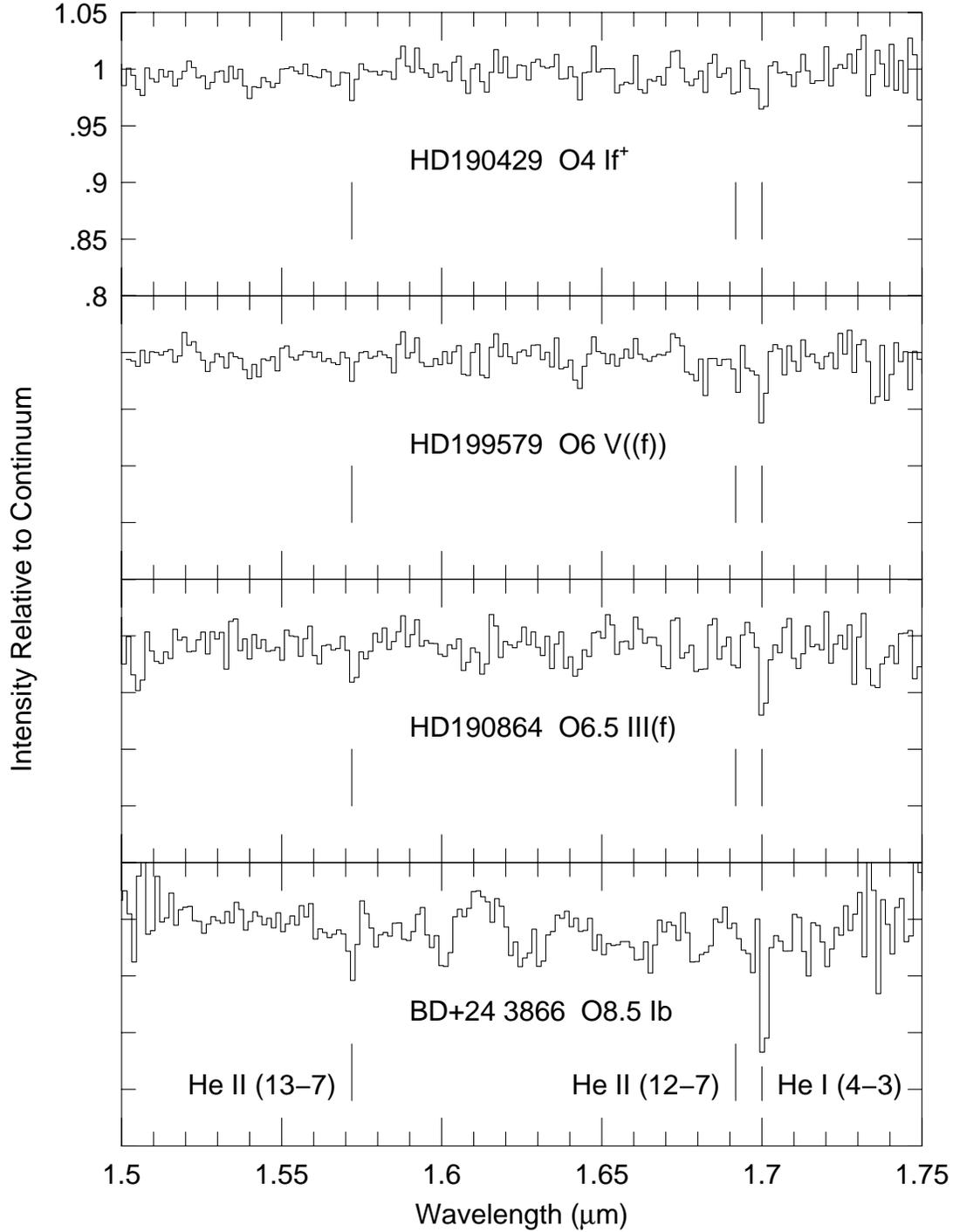}{7 in}{0}{80}{80}{-260}{-40}
\caption[]{Detail of $H-$band spectra for the O stars (same spectra as
plotted in Figure~\ref{spect}) showing probable \ion{He}{2} absorption
at 1.5719 \mic. \ion{He}{2} 1.6918 \mic \ absorption is marked although not
clearly detected; see text. The intensity scale is the same for each
star.
}
\label{heii}
\end{figure}

Figure~\ref{heii} shows a more detailed look at the $H-$band spectra of
the four O stars. The figure suggests a tentative detection of
\ion{He}{2}
(13-7) at 1.5719 \mic \ and perhaps \ion{He}{2} (12--7) at 1.6918 \mic. The
former is clearly in a region of higher signal--to--noise; the latter
sits in a region with larger telluric features (the strongest of which
are to the blue of the \ion{He}{1} 1.7002 \mic \ line).
The 1.5719 \mic \ line appears in both of the O star spectra of
Lan\c{c}on \& Rocca--Volmerange (1992\markcite{lv92})
as well, including their spectrum
of HD 199579, one of the stars observed in our sample.
Detection of \ion{He}{2} is difficult since the 1.5719 \mic \ line is near
the Brackett 15--4 line (1.5701 \mic); and hence, the
\ion{He}{2} line could depend on how well we have corrected the Brackett
series lines in the A star which was used as a atmospheric standard (see
\S2). The \ion{He}{2} lines may prove more useful when observed at higher
spectral resolution.

\subsection{$I'$ and $J-$band Spectra}

\begin{figure}
\plotfiddle{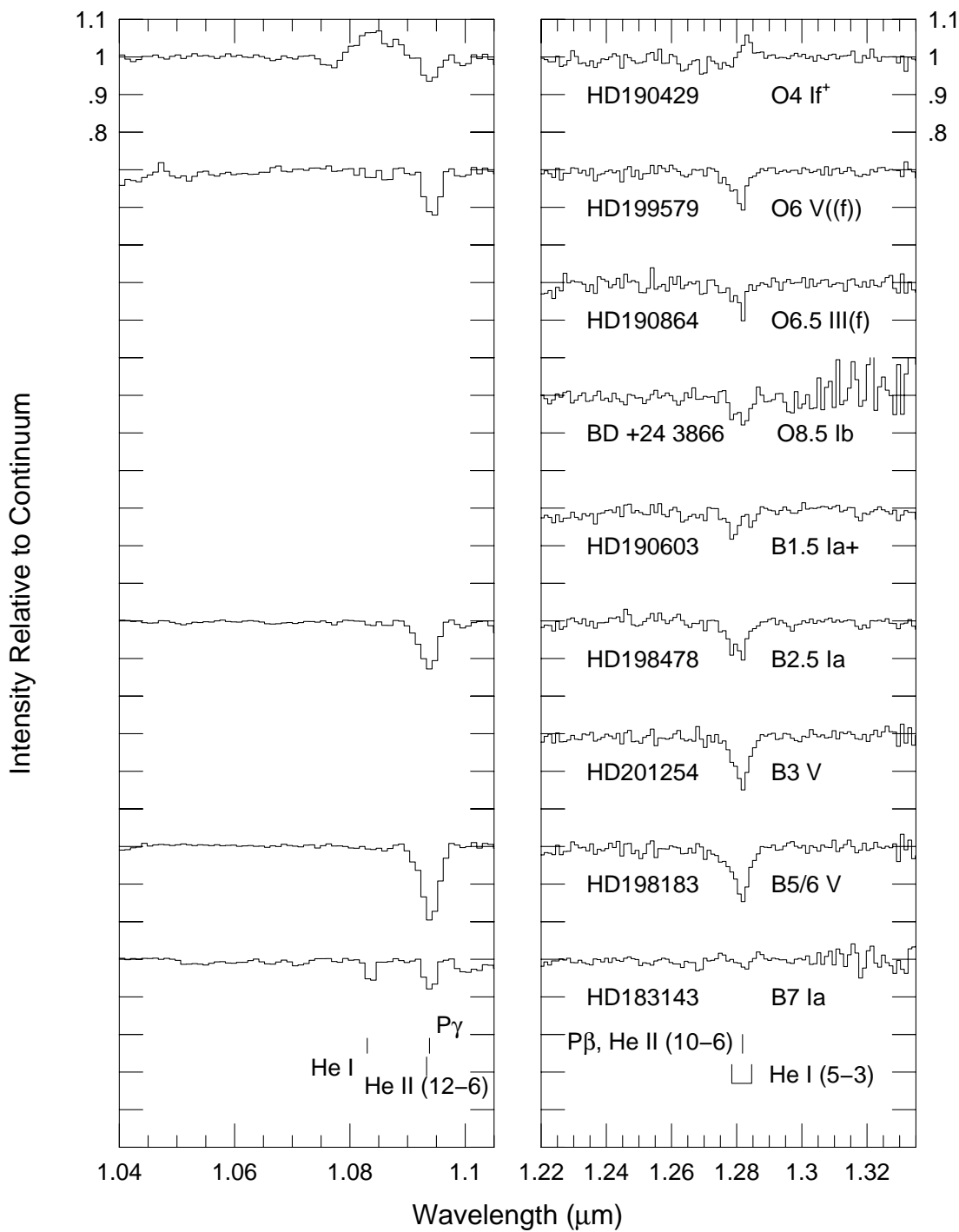}{7 in}{0}{80}{80}{-260}{-40}
\caption[]{$\lambda / \Delta\lambda \approx 570$ $I'$ and $J-$band
spectra of OB stars. The
spectra have been normalized by a low order fit to the continuum.
Pa$\gamma$ (1.0938 \mic)  and Pa$\beta$ (1.2818 \mic) may have
contributions due to \ion{He}{1} and \ion{He}{2}. The intensity scale is
the same for each star.
}
\label{spect2}
\end{figure}

Figure~\ref{spect2} shows spectra of the OB stars near 1.1 and 1.3 \mic.
The primary lines which fall within our $I'$ and $J-$band spectra are
due to H and \ion{He}{1}, as in the $H-$band spectra. However the H lines,
Pa$\gamma$ (1.0938 \mic) and Pa$\beta$ (1.2818 \mic)
are blended with lines of \ion{He}{1}
and \ion{He}{2} at our resolution (see Figure~\ref{spect2}),
so they are of limited use for determining spectral types.
We note that \ion{He}{1} 1.0830 \mic \ is present in several of the stars
(HD190429, P Cygni profile; HD183143).
We anticipate that the behavior of this triplet line
(2$p ^3P^{\circ}$--2$s ^3S$) will $not$ be
well correlated with spectral type, following the example of the
analogous singlet \ion{He}{1} feature at 2.0581 \mic \ (HCR96).
We have only five $I'$ spectra in our initial sample, so we do not attempt
to make any firm conclusions about this wavelength region.
In any case,
the $I'$ and $J-$ bands should be useful for comparison to model atmosphere
predictions, especially when they are obtained for ``free,'' i.e. in
cross--dispersed mode.

\section{SUMMARY}

We have presented $H-$band spectra of a preliminary set of OB stars
and identified absorption features that can be used to classify young,
massive stars in obscured H II regions and/or along sight lines with
large optical extinction due to interstellar dust. The absorption
strength of the \ion{He}{1} (4--3) 1.7002 \mic \ line, in conjunction with
the
presence or absence of the $H-$band lines of the Brackett series,
is well correlated with optical spectral types
and can be used as a coarse spectral classification. The behavior of the
infrared lines is analogous to well--studied lines in the optical
wavelength region including basic effects due to excitation
(temperature) and gravity (pressure). In particular, we see
the effects of linear Stark broadening, resulting from changes in
pressure in B dwarfs and supergiants, on the Brackett lines which
will be useful as a luminosity indicator in the B stars.
We have also presented spectra of the OB stars between
1.1 and 1.3 \mic. Features in these wavelength regions may prove useful
in cases where excess emission is still strong at $H$ or for the
purposes of testing atmosphere models of hot stars.

Support for this work was provided by NASA through grant number
HF 01067.01 -- 94A from the Space Telescope Science Institute, which is
operated by the Association of Universities for Research in Astronomy,
Inc., under NASA contract NAS5--26555. OSIRIS was built with support
from NSF grants AST 90--16112 and AST 9218449.
This work was also supported by NSF grant AST 93--14808.


\newpage


\end{document}